\documentclass[epsfig,twocolumn,preprintnumbers,amsmath,amssymb]{revtex4}
\usepackage{epsfig}

\begin{document}

\title{
Energy dependence of directed flow over a wide range of
pseudorapidity in Au+Au collisions at RHIC}
\author {
B.B.Back$^1$, M.D.Baker$^2$, M.Ballintijn$^4$, D.S.Barton$^2$,
R.R.Betts$^6$, A.A.Bickley$^7$, R.Bindel$^7$, A.Budzanowski$^3$,
W.Busza$^4$, A.Carroll$^2$, Z.Chai$^2$, M.P.Decowski$^4$,
E.Garc\'{\i}a$^6$, T.Gburek$^3$, N.George$^{1,2}$,
K.Gulbrandsen$^4$, S.Gushue$^2$, C.Halliwell$^6$, J.Hamblen$^8$,
M.Hauer$^2$, G.A.Heintzelman$^2$, C.Henderson$^4$, D.J.Hofman$^6$,
R.S.Hollis$^6$, R.Ho\l y\'{n}ski$^3$, B.Holzman$^2$,
A.Iordanova$^6$, E.Johnson$^8$, J.L.Kane$^4$, J.Katzy$^{4,6}$,
N.Khan$^8$, W.Kucewicz$^6$, P.Kulinich$^4$, C.M.Kuo$^5$,
W.T.Lin$^5$, S.Manly$^8$, D.McLeod$^6$, A.C.Mignerey$^7$,
R.Nouicer$^{2,6}$, A.Olszewski$^3$, R.Pak$^2$, I.C.Park$^8$,
H.Pernegger$^4$, C.Reed$^4$, L.P.Remsberg$^2$, M.Reuter$^6$,
C.Roland$^4$, G.Roland$^4$, L.Rosenberg$^4$, J.Sagerer$^6$,
P.Sarin$^4$, P.Sawicki$^3$, H.Seals$^2$, I.Sedykh$^2$,
W.Skulski$^8$, C.E.Smith$^6$, M.A.Stankiewicz$^2$, P.Steinberg$^2$,
G.S.F.Stephans$^4$, A.Sukhanov$^2$, J.-L.Tang$^5$, M.B.Tonjes$^7$,
A.Trzupek$^3$, C.Vale$^4$, G.J.van~Nieuwenhuizen$^4$,
S.S.Vaurynovich$^4$, R.Verdier$^4$, G.I.Veres$^4$, E.Wenger$^4$,
F.L.H.Wolfs$^8$, B.Wosiek$^3$, K.Wo\'{z}niak$^3$, A.H.Wuosmaa$^1$,
B.Wys\l ouch$^4$\\
$^1$ Physics Division, Argonne National Laboratory, Argonne, IL
60439-4843\\ $^2$ Chemistry and C-A Departments, Brookhaven
National Laboratory, Upton, NY 11973-5000\\ $^3$ Institute of
Nuclear Physics PAN, Krak\'{o}w, Poland\\ $^4$ Laboratory for
Nuclear Science, Massachusetts Institute of Technology, Cambridge,
MA 02139-4307\\ $^5$ Department of Physics, National Central
University, Chung-Li, Taiwan\\ $^6$ Department of Physics,
University of Illinois at Chicago, Chicago, IL 60607-7059\\ $^7$
Department of Chemistry and Biochemistry, University of Maryland,
College Park, MD 20742\\ $^8$ Department of Physics and Astronomy,
University of Rochester, Rochester, NY 14627\\ }
\date{\today}

\begin{abstract}\noindent
We report on measurements of directed flow as a function of
pseudorapidity in Au+Au collisions at energies of $\sqrt{s_{_{NN}}}
=$ 19.6, 62.4, 130 and 200 GeV as measured by the PHOBOS detector at
the Relativistic Heavy Ion Collider (RHIC).  These results are
particularly valuable because of the extensive, continuous
pseudorapidity coverage of the PHOBOS detector.  There is no
significant indication of structure near midrapidity and the data
surprisingly exhibit extended longitudinal scaling similar to that
seen for elliptic flow and charged particle pseudorapidity density.
\end{abstract}

\maketitle


The study of collective flow in ultrarelativistic nuclear collisions
provides insight into the equation of state, degree of
thermalization, and the early stages of the hot, dense matter
created.  The elliptic flow parameter, $v_{2}$, has been studied
extensively over a wide range of collision energies and
pseudorapidity~\cite{phwp,stphwp}. The directed flow parameter,
$v_{1}$, however, has been studied in less detail at RHIC
energies~\cite{starv1v4,starflow,star62v1}.

The PHOBOS detector is composed of several subsystems (see
Ref.~\cite{phobos_det} for details). The most important for this
analysis were the silicon multiplicity array, which consists of an
octagonal multiplicity detector (OCT), a silicon vertex detector
(VTX), and three annular ring multiplicity detectors (RINGS) located
on each side of the collision point. PHOBOS has the ability to
measure nearly all charged particles due to the $\sim4\pi$ solid
angle coverage and to record particles with transverse momenta down
to about 35~MeV/c (140~MeV/c) for pions (protons) at $\eta=0$ and
4~MeV/c (10~MeV/c) for $\eta\sim$ 4--5.

This analysis is based on data sets for Au+Au collisions at
$\sqrt{s_{_{NN}}} =$ 19.6, 62.4, 130, and 200 GeV, as used in the
elliptic flow study~\cite{v2limfrag}. All data sets were taken with
the spectrometer magnetic field off, except for data taken at 130
GeV, where field-on data was included to maximize statistics.
Details on event selection and signal processing can be found
in~\cite{phwp,phobos_det}. The results are shown in the most central
40\% of the total inelastic cross-section for which the trigger
system was fully efficient at all four energies. More information on
triggering and centrality determination can be found in~\cite{phwp}.
Monte Carlo simulations of the detector performance were based on
the Hijing event generator~\cite{hijing} and the
GEANT~3.211~\cite{geant} simulation package, folding in the signal
response for scintillator counters and silicon sensors.

The analysis is based on the anisotropy of the azimuthal
distribution of charged particles detected in the silicon pads of
the PHOBOS multiplicity array. The analysis uses a subevent
technique where hits produced in one area of the detector are
correlated with an event plane angle found from hits in another
region~\cite{pandv}.

% figure 1, original width 16.0cm
\begin{figure*}[t]
%\begin{figure}[h]
\centerline{\epsfig{file=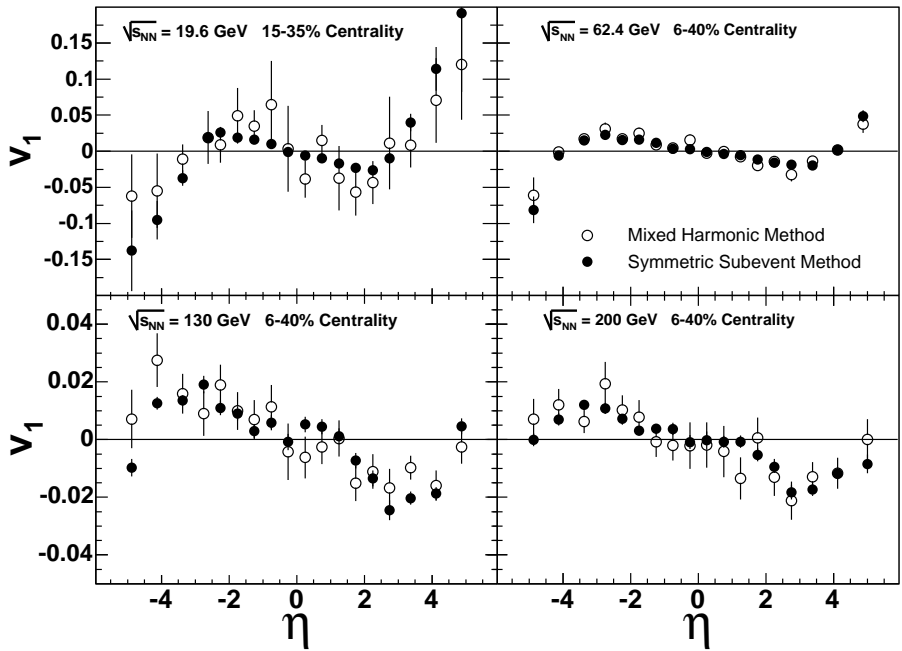,width=12.0cm}} \caption{ Measured
directed flow as a function of $\eta$ in Au+Au collisions at
$\sqrt{s_{_{NN}}} =$ 19.6, 62.4, 130, and 200 GeV, using the mixed
harmonic event plane method (open points) overlayed with the
standard symmetric $\eta$ subevent method (closed points). Note the
different vertical axis scales between the upper and lower panels.
The centrality ranges shown for both methods are those which give
good mixed harmonic reaction plane sensitivity. For clarity only the
statistical errors are shown.} \label{v1symmixedcomp}
%\end{figure}
\end{figure*}
% end figure 1

Directed flow is quantified by measuring the first harmonic,
$v_{1}$, of the Fourier decomposition of the particle azimuthal
angle distribution,

\begin{equation}
 \frac{dN}{d(\phi-\psi_{R})}
= {1\over 2 \pi } \big( 1 + \sum_{n=1}^{\infty} 2{\rm
v}_{n}\cos[n(\phi-\psi_{R})] \big),
\end{equation}
where $\psi_{R}$ is the true reaction plane angle defined by the
impact parameter and beam axis.

Measuring directed flow using a subevent technique is associated
with several pitfalls because global momentum conservation can
produce non-flow correlations between the subevent and the particle
under study~\cite{flowmomcons}.  This analysis circumvents this
correlation by using subevent windows that are symmetric about
midrapidity~\cite{BDO}, thereby canceling the back-to-back momentum
conservation recoil because each subevent is composed of two equal
sections in the negative and positive $\eta$ hemispheres. The
subevent regions used in the event plane calculations are located in
the OCT ($1.5<|\eta|<3$) and RINGS ($3<|\eta|<5$) subdetectors. The
OCT subevent is used to find $v_{1}$ in the RINGS region
($|\eta|>3$). Likewise, the RINGS subevent is used to find $v_{1}$
in the OCT ($|\eta|<3$) region. For each $\eta$-symmetric subevent
window, a resolution correction is applied that was equal to
\begin{equation}
\frac{1}{\sqrt{2 <cos (\psi_{1N}-\psi_{1P})>}},
\end{equation}
where the N and P labels denote event planes found in the negative
and positive halves of each subevent window. The centrality averaged
resolution correction for the octagon subevent was 4.1 for 19.6 GeV
and 3.5 for the other three energies. The centrality averaged ring
subevent resolution correction was 1.9, 3.3, 4.5, and 3.8 for the
19.6, 62.4, 130 ,and 200 GeV data sets, respectively.

This analysis accepted collisions within $\pm10$ cm of the nominal
vertex position.  In this detector region there exist holes in the
octagon acceptance to avoid shadowing the vertex and spectrometer
detectors. The detector was symmetrized using the procedure
described in~\cite{v2limfrag}.

% figure 2
\begin{figure*}[t]
%\begin{figure}[h]
%\centerline{ \epsfig{file=v1vseta.eps,width=9.0cm} used to be 18.2 cm
\centerline{ \epsfig{file=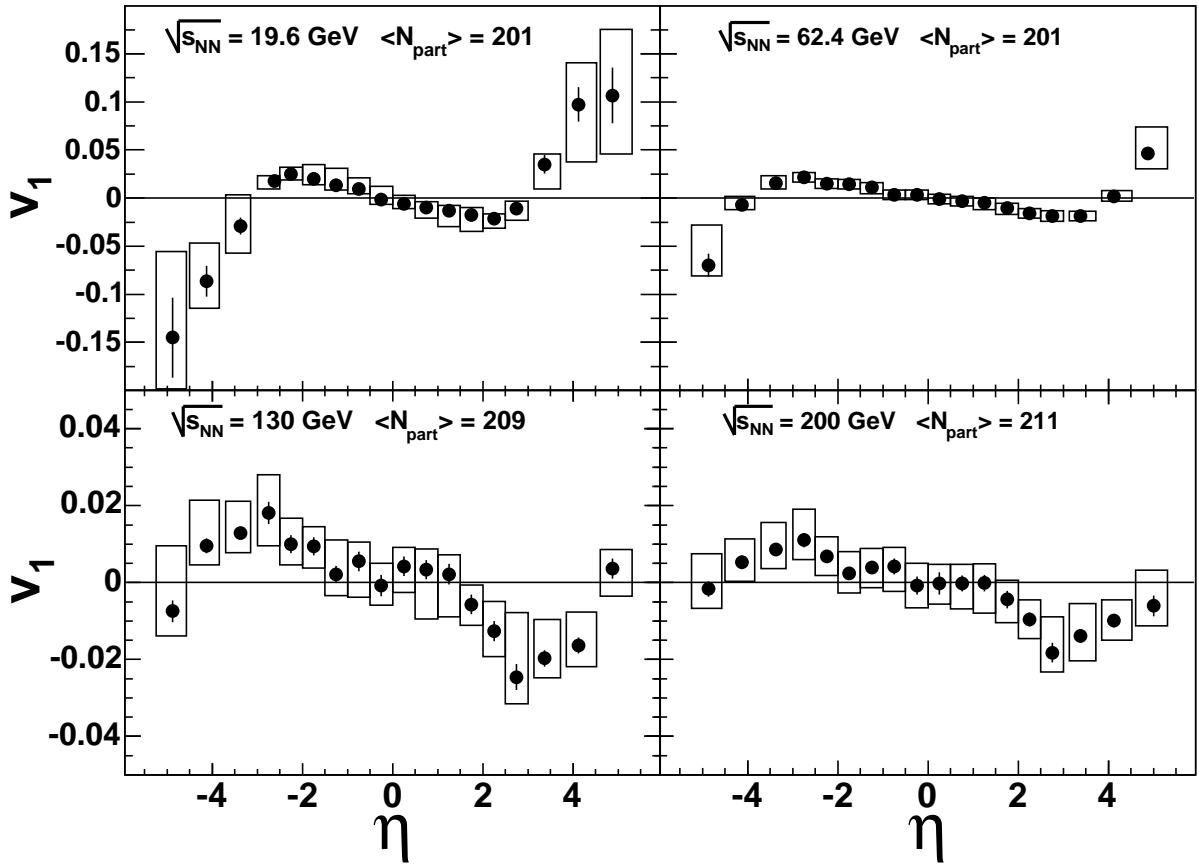,width=16.0cm} } \caption{
Directed flow of charged particles in Au+Au collisions as a function
of $\eta$, averaged over centrality (0--40\%), shown separately for
four beam energies. Note the different vertical axis scales between
the upper and lower panels. The boxes represent systematic
uncertainties at 90\% C.L., and $\langle N_{part} \rangle$ gives the
average number of participants for each data sample.}
\label{v1stacked}
%\end{figure}
\end{figure*}
% end figure 2

In addition, weights were applied to the pads in the symmetrized
detector hit map to correct for phase space differences between the
detector pads, as well as to account for instances where more than
one track passes through a pad.  The weighting procedure is the same
as applied in the analysis of elliptic flow~\cite{v2limfrag}.

Monte Carlo studies showed a suppression of the reconstructed flow
signal that is dominated by background particles that do not carry
flow information, as well as the loss of sensitivity due to the hit
map symmetrization and the occupancy correction algorithm. As in the
elliptic flow analysis~\cite{v2limfrag}, this suppression is
corrected using the ratio of reconstructed to input flow from the
simulation.  Typical correction levels were in the 25-30\% range for
the results shown.

In the subevent method described above, while the flow itself is
measured using symmetric subevents, the resolution correction
correlates portions of the detector that lie in the forward and
backward $\eta$ regions. Thus, it is possible that a small non-flow
correlation due to momentum conservation affects the final result
through the resolution correction. In order to estimate the
potential size of this and any other non-flow correlations
contributing to the signal, we also analyzed the data using a mixed
harmonic event plane analysis~\cite{BDO2}. In the mixed harmonic
analysis, the reaction plane, $\psi_{2}$, is determined using the
elliptic flow information and the directed flow signal perpendicular
to $\psi_{2}$ (out-of-plane) is subtracted from that which is in the
plane of $\psi_{2}$ (in-plane).  Since the true directed flow signal
is in-plane, the assumption is that the directed flow signal
out-of-plane is due to non-flow correlations.

Specifically, in our implementation of the mixed harmonic analysis,
$\psi_{2}$ was found in two subevents from $-3<\eta<-0.1$ and
$0.1<\eta<3$ and used along with the $\eta$-symmetric $\psi_{1}$
event planes defined above to find $v_{1}\{\psi_{1},\psi_{2}\}$ as
outlined in~\cite{starflow}. Two $\psi_{1}$ event plane angles and
two $\psi_{2}$ angles were necessary in order to find $v_{1}$ in all
regions of pseudorapidity such that the particle under study did not
fall into the regions where either $\psi_{1}$ or $\psi_{2}$ event
plane angles were determined.

Fig.~\ref{v1symmixedcomp} shows the fully corrected signal for the
directed flow at all energies as a function of pseudorapidity for
both analysis methods.  The 1$\sigma$ statistical errors are shown
as solid bars. In both methods, the statistical errors exhibit a
point-to-point correlation due to shared event plane and event plane
resolution determination. The mixed harmonic method gives results
which are consistent with the symmetric subevent method at 62.4,
130, and 200 GeV. At 19.6 GeV the mixed harmonic analysis results
are in reasonable agreement with the symmetric subevent method;
however, the analyzing power of the mixed harmonic method is
diminished at this energy due to the weak elliptic flow signal, as
well as a very small event sample.

The agreement between these methods implies that the reaction plane
determined by elliptic flow is the same as that determined by
directed flow, within errors. This in turn means that the flow and
the reaction plane that we see in Au+Au collisions is dominated by a
global flow of the particles with minimal effects from ``non-flow
correlations''. Furthermore, since the $v_2$ reaction plane is
dominated by $\eta$ near zero and $v_1$ by high $|\eta|$, this
result indicates that the reaction plane orientation is consistent
over the entire pseudorapidity range.

Fig.~\ref{v1stacked} shows the results from the symmetric subevent
method with the 90\% C.L. systematic errors. Several aspects of the
analysis were studied in order to establish the systematic errors.
These include hit definition, hit merging, subevent definition,
knowledge of the beam orbit relative to the detector, $dN/d\eta$
distribution, hole filling procedure, consistency of $v_{1}$ result
when rotated by 180 degrees, magnetic field configuration and the
suppression correction determination. The systematic error from each
source was estimated by varying that specific aspect within
reasonable limits and quantifying the change in the final $v_{1}$ as
a function of $\eta$. Also, the difference between the results from
the symmetric subevent method and an odd-order polynomial fit to the
mixed harmonic method was included in the systematic error. The
individual contributions were added in quadrature to obtain the
final systematic errors.

Historically, $v_{1}$ has been defined to be positive (negative) at
high positive (negative) $\eta$ where spectator matter is thought to
dominate the signal~\cite{na49flow}.  We have preserved that
convention here, although it is important to note that the spectator
region falls outside of our acceptance at the higher energies.
Consequently, the regions of $\eta$ used to find the direction of
$\psi_{1}$ have varying spectator content as the collision energy
increases. Thus, it is necessary to invert the sign of $v_{1}$ at
130 and 200 GeV in order to preserve the sign convention from the
lower energies and make a direct comparison of the shapes as a
function of energy, as shown in Fig.~\ref{v1stacked}.

The results in Fig.~\ref{v1stacked} show the evolution of $v_{1}$ as
the collision energy increases.  All four energies exhibit a $v_{1}$
signal passing smoothly through zero at $\eta=0$ as expected,
indicating that there are no momentum conservation biases in the
data. The $v_{1}$ becomes more negative with $\eta$ at each energy,
until a ``turnover'' point is reached, and the $v_{1}$ from both
19.6 and 62.4 GeV becomes positive at very high pseudorapidities.
This turnover at all energies and the large signal seen at high
$|\eta|$ for the lower energies are features uniquely observed by
PHOBOS. These effects may be due to protons and nuclear fragments
taking over from pions as the dominant contributors to the directed
flow signal at high $|\eta|$.

% figure 3
\begin{figure}[h]
\vspace*{0.3cm} \centerline{ \epsfig{file=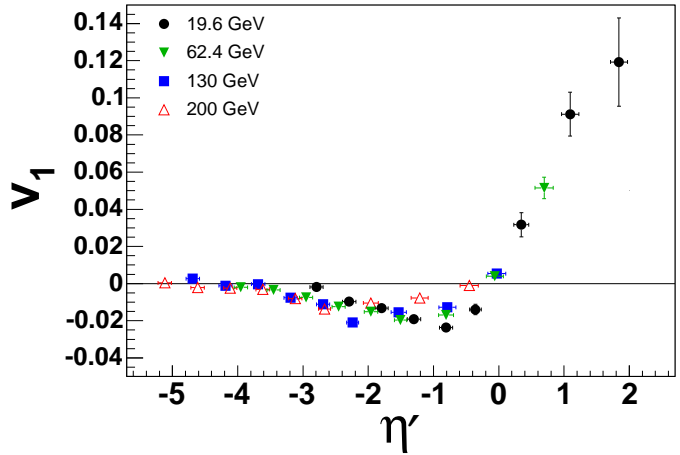,width=9.0cm} }
 \caption{
Directed flow, averaged over centrality (0--40\%), as a function of
$\eta'=|\eta|-y_{beam}$ for four beam energies. The error bars
represent the 1$\sigma$ statistical errors only.}
\label{v1etaprimefold}
\end{figure}

The results at 62.4 and 200 GeV are in qualitative agreement with
results from STAR~\cite{starflow,star62v1}.  Both experiments show
$v_{1}\sim$ 0 for an extended region about midrapidity at 200 GeV,
while $|v_{1}|$ increases as $|\eta|$ increases.  At 62.4 GeV,
PHOBOS observes a turnover of the $v_{1}$ signal that occurs at
smaller pseudorapidity than what is reported in the STAR data. This
may indicate that $v_1$ at high $|\eta|$ is sensitive to the
transverse momentum range included in the measurement. Recall that
PHOBOS measures protons down to $p_{_{T}} \sim 10$~MeV/c while STAR
has a cutoff at $p_{_{T}}\sim 150$~MeV/c.

Fig.~\ref{v1etaprimefold} shows the directed flow where data points
from the positive and negative $\eta$ regions have been averaged
together and plotted as a function of $\eta'=|\eta|-y_{beam}$. Since
the directed flow curves are odd functions, the negative $\eta$
region was multiplied by -1 before the averaging was performed to
avoid cancelation. Within the systematic errors (shown in
Fig.~\ref{v1stacked}), it appears that all curves scale throughout
the entire region of $\eta'$ overlap, showing that, within errors,
the directed flow exhibits the longitudinal scaling behavior already
observed in the elliptic flow~\cite{v2limfrag} and charged particle
multiplicity~\cite{limfrag}. This confirms and expands on an earlier
observation of this scaling in the directed flow between RHIC and
SPS results~\cite{star62v1}.

In summary, the pseudorapidity dependence of directed flow has been
measured for several collision energies.  At each energy, the
$v_{1}$ signal is small at midrapidity and grows with increasing
$|\eta|$. At very high $|\eta|$, a turnover of $v_{1}$ is observed,
possibly due to protons and nuclear fragments dominating the flow
signal in this range. When studied as a function of $\eta'$, $v_{1}$
appears to scale throughout the entire $\eta'$ overlap region at all
energies.

Acknowledgements:
%We acknowledge the generous support of the
%Collider-Accelerator Department
% (including RHIC project personnel) and Chemistry Departments at BNL.  We
% thank Fermilab and CERN for help in silicon detector assembly.  We thank the
% MIT School of Science and LNS for financial support.
%
This work was partially supported by U.S. DOE grants
DE-AC02-98CH10886, DE-FG02-93ER40802,
DE-FC02-94ER40818,  % MIT
DE-FG02-94ER40865, DE-FG02-99ER41099, and W-31-109-ENG-38, by U.S.
NSF grants 9603486, % Phobos TOF
0072204,            % Rochester until 6/03
and 0245011,        % Rochester starting 6/03
by Polish KBN grant 1-P03B-06227, and by NSC of Taiwan Contract NSC
89-2112-M-008-024.


\begin{thebibliography}{99}
%
%
\bibitem{phwp}
%white papers (phobos)
B.B.~Back {\it et al.}, Nucl. Phys. {\bf A757}, 28 (2005).
%
\bibitem{stphwp}
%white papers (star, phenix)
J.~Adams {\it et al.}, Nucl. Phys. {\bf A757}, 102 (2005); K.~Adcox
{\it et al.}, Nucl. Phys. {\bf A757}, 184 (2005).
%
\bibitem{starv1v4}
%STAR Collaboration,
J.~Adams {\it et al.}, Phys. Rev. Lett. {\bf 92}, 062301 (2004).
%
\bibitem{starflow}
%STAR Collaboration, nucl-ex/0409033
J.~Adams {\it et al.},  Phys. Rev. C {\bf72}, 014904 (2005).
%
\bibitem{star62v1}
%Star Collaboration, nucl-ex/0409029
J.~Adams {\it et al.}, submitted to Phys. Rev. C (Rapid Comm.)
[nucl-ex/0510053].
% A.H. Tang {\it et al.}, J. Phys. {\bf G31}, S35-S40 (2005) [nucl-ex/0409029].
%
\bibitem{phobos_det}
%PHOBOS Collaboration,
B.B.~Back {\it et al.},  Nucl. Instrum. Meth. {\bf A 499}, 603
(2003).
%
\bibitem{v2limfrag}
%PHOBOS Collaboration,
B.B.~Back {\it et al.}, Phys. Rev. Lett. {\bf 94}, 122303 (2005).
%
\bibitem{hijing} X.N.Wang and M.Gyulassy, Phys. Rev. D
{\bf44},  3501 (1991).  We used standard Hijing v1.35.
%
\bibitem{geant} GEANT 3.211, CERN Program Library.
%
\bibitem{pandv}  A.M. Poskanzer and S.A. Voloshin, Phys. Rev. C {\bf
58}, 1671 (1998).
%
\bibitem{flowmomcons}  N. Borghini, P.M. Dinh, J.-Y. Ollitrault, A.M. Poskanzer and S.A. Voloshin, Phys. Rev. C {\bf66}, 014901 (2002).
%
\bibitem{BDO}  N. Borghini, P.M. Dinh, and J.-Y. Ollitrault, Phys. Rev. C {\bf62}, 034902 (2000).
%
\bibitem{BDO2}
N. Borghini, P.M. Dinh, and J.-Y. Ollitrault, Phys. Rev. C {\bf66},
014905 (2002).
%
\bibitem{na49flow} C. Alt {\it et al.}, Phys. Rev. C {\bf68}, 034903
(2003).
%
\bibitem{limfrag}
%PHOBOS Collaboration,
B.B.~Back {\it et al.},  Phys. Rev. Lett. {\bf 91}, 052303 (2003).
%
\end{thebibliography}
\end{document}